\documentclass[conference, 10pt]{IEEEtran}
\usepackage{cite}
\usepackage{graphicx}
\usepackage{algorithm,algorithmic}
\usepackage[caption=false, font=footnotesize]{subfig}
\usepackage{url}

\begin{document}
\title{Improving Device-Edge Cooperative Inference of Deep Learning via 2-Step Pruning}

\author{\IEEEauthorblockN{Wenqi Shi\IEEEauthorrefmark{1}, Yunzhong Hou\IEEEauthorrefmark{1}, Sheng Zhou\IEEEauthorrefmark{1}, Zhisheng Niu\IEEEauthorrefmark{1},
Yang Zhang\IEEEauthorrefmark{2} and Lu Geng\IEEEauthorrefmark{2}}
        \IEEEauthorblockA{\IEEEauthorrefmark{1}Beijing National Research Center for Information Science and Technology,\\Department of Electronic Engineering, Tsinghua University, Beijing 100084, China\\
        	Email: swq17@mails.tsinghua.edu.cn, Hou\_Yz@outlook.com, \{sheng.zhou,niuzhs\}@tsinghua.edu.cn}

		\IEEEauthorblockA{\IEEEauthorrefmark{2}Hitachi (China) Research \& Development Cooperation, Beijing 100190, China\\     
		Email: \{zhangyang, lgeng\}@hitachi.cn}}

\maketitle

\begin{abstract}
Deep neural networks (DNNs) are state-of-the-art solutions for many machine learning applications, and have been widely used on mobile devices.
Running DNNs on resource-constrained mobile devices often requires the help from edge servers via computation offloading.
However, offloading through a bandwidth-limited wireless link is non-trivial due to the tight interplay between the computation resources on mobile
devices and wireless resources.
Existing studies have focused on cooperative inference where DNN models are partitioned at different neural network layers,
and the two parts are executed at the mobile device and the edge server, respectively.
Since the output data size of a DNN layer can be larger than that of the raw data, 
offloading intermediate data between layers can suffer from high transmission latency under limited wireless bandwidth.
In this paper, we propose an efficient and flexible 2-step pruning framework for DNN partition between mobile devices and edge servers. 
In our framework, the DNN model only needs to be pruned once in the training phase where unimportant convolutional filters are removed iteratively.
By limiting the pruning region, our framework can greatly reduce either the wireless transmission workload of the device or the total computation workload.
A series of pruned models are generated in the training phase, 
from which the framework can automatically select to satisfy varying latency and accuracy requirements.
Furthermore, coding for the intermediate data is added to provide extra transmission workload reduction.
Our experiments show that the proposed framework can achieve up to 25.6$\times$ reduction on transmission workload, 6.01$\times$ acceleration on total computation 
and 4.81 $\times$ reduction on end-to-end latency as compared to partitioning the original DNN model without pruning.
\end{abstract}

\IEEEpeerreviewmaketitle

\section{Introduction}
In recent years, deep neural network (DNN) has become the most influential model of machine learning, 
and has achieved great successes in a wide range of applications\cite{goodfellow2016deep}.
Advanced DNN models have brought the possibility of intelligent end-user applications on mobile devices, such as intelligent personal assistants (IPAs), 
smart wearable devices and autonomous vehicles. 

However, the timely and reliable inference of DNNs require powerful computation resources.
Since the mobile devices usually have much weaker computing capability and limited energy supply (typically battery) compared to the edge servers,
completing DNN inference on mobile devices within a reasonable latency and energy consumption is challenging.
To solve this problem, the wisdom from mobile edge computing suggests to host all computation on edge servers\cite{abbas2018mobile}.
In this case, the raw data collected by the mobile device is sent to the edge server via the wireless channel.
The edge server sends the inference result back to the mobile device after the DNN computation is completed.
Uploading raw data (e.g., images, audios, videos) is bandwidth consuming and will introduce unpredictable latency due to the wireless channel fading.

\begin{figure}[!t]
  \centering
  \includegraphics[width=3.2in]{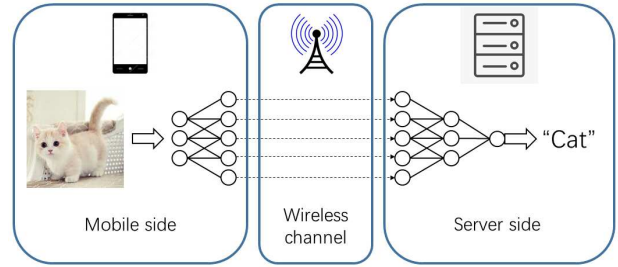}
  \caption{An illustration of device-edge cooperative inference.}
  \label{device-edge}
  \end{figure}

Beyond mobile-only and edge-only approaches, 
partitioning the DNN model computation between the mobile devices and the edge servers has advantages on balancing the transmission and
computation workload between mobile devices and servers, so as to optimize the end-to-end DNN completion latency and the energy consumption of mobile devices.
Some researchers have put efforts on the layer-level partition of DNN\cite{Neurosurgeon,teerapittayanon2017distributed, eshratifar2018jointdnn, li2018edge}.
As illustrated in Fig.~\ref{device-edge}, the mobile device computes the inference up to an intermediate layer of the DNN model, 
and transmits the output features of that layer to the server for completing the rest of the DNN.
The choice of the best partition point depends not only on the various system factors
(e.g., wireless channel state, wireless communication technique, computation capability of mobile devices and edge servers), but also the DNN model.
Since the data size increases after convolution layers, the data volume of the output features from the intermediate layer can be much larger than the raw data,
and uploading intermediate data will suffer from even higher transmission latency and energy consumption compared to uploading the raw data.
As a result, a fixed DNN model can hardly benifit from partitioning under many circumstances\cite{Neurosurgeon}.
In addition, early exit\cite{li2018edge} and feature coding\cite{ko2018edge} techniques
are proposed to reduce the volume of the intermediate data to be transmitted.
The main drawback of the feature coding is the extra computing latency introduced by coding,
while the early exit suffers from low accuracy under severe latency constraints\cite{li2018edge}.
Therefore there is an emerging need for a well-designed framework that can fine-tune the DNN model
in order to optimize not only the partition point selection but also the computation/transmission workload reductions.

In this paper, we propose a 2-step pruning framework which combines pruning method and DNN partition in order to enhance the end-to-end latency in device-edge cooperative inference.
Inspired by the fact that the data volume of the output feature from each intermediate layer is determined by the product of the length,
height and number of feature maps (or channels),
we apply a state-of-the-art channel pruning method to prune unimportant feature maps in each layer.
The conventional pruning process is split into two pruning steps, where the first step is mainly for the computation workload reduction
and the second step is mainly for the transmission workload reduction.
Our framework generates a series of pruned DNN models with different compression ratios via the two pruning steps,
and can automatically choose the best pruned DNN model and corresponding partition point according to the system factors (e.g., latency requirements and accuracy requirements).
Experiments show that our 2-step pruning framework can greatly improve the end-to-end latency and maintain higher accuracy under
limited bandwidth constraints in most circumstances compared to existing feature coding approach.

The rest of the paper is organized as follows. 
In section II, we review the background and related work.
In section III, the proposed 2-step pruning framework is introduced. 
The experiments are presented and discussed in section IV.
Finally, the paper is concluded in section V.

\section{Background and Related Work}

DNNs are composed of a series of layers and each layer is comprised of a collection of processing elements (neurons).
CNN is a special class of DNN where at least one layer employs convolution operation.
This convolution layer convolve the input feature maps with a set of learned filters to generate the output feature maps.
CNNs are used extensively in computer vision and natural language processing applications, including image classification,
object detection, video recognition and document classification.
Most existing CNN models are computationally expensive thus efficient DNN inference on resource-constrained end-devices has received many research efforts.
There are two major fields under investigation.

\textbf{Optimizing Deep Learning Model:}
On the software side, some researchers have proposed deep network models that are much smaller than normal ones without 
sacrificing too much accuracy\cite{howard2017mobilenets}. 
Some others have focused on model compression techniques
that reduce the redundancy in the original model to get an efficient model\cite{molchanov2016pruning, luo2017thinet, he2017channel}. 
In NVIDIA’s channel pruning work\cite{molchanov2016pruning}, entire feature maps from the output of convolutional layer are pruned. 
The resulting network can run efficiently and the size of intermediate data between two layers can be reduced as a byproduct. 
On the hardware side, mobile devices and edge servers can embed deep learning inference
engine to enhance the latency and energy efficiency with the help of architectural acceleration techniques\cite{ovtcharov2015accelerating, han2016eie}.

\textbf{Mobile Device and Edge Server Cooperation:}
Some recent studies have proposed distributed deep neural network over mobile devices and edge servers. 
In\cite{Neurosurgeon}, Kang et al. have investigated layer level partitions of DNN models with 3 typical wireless communication technologies: 3G, LTE and WiFi. 
The results show that a portion of DNN models can benefit from partitioning while others will suffer from high transmission latency in transmitting intermediate data. 
Eshratifar et al. formulate the DNN partition problem into a shortest path problem and use an approximation solution to solve the problem\cite{eshratifar2018jointdnn}. 
Furthermore, they use PNG coding to reduce data volume of the intermediate data. 
Later, Ko et al. examine both the lossy and the lossless JPEG coding for the intermediate data of partitioned DNNs\cite{ko2018edge}.
Their results show that coding can greatly improve the energy efficiency and computation throughput with little accuracy loss. 
A modified DNN structure\cite{teerapittayanon2017distributed, li2018edge}, where early exit network branches are added to the original network, has been proposed for the device-edge synergy.
Their evaluations demonstrate the effectiveness of the modified DNN structure in enabling low-latency inference with tolerable accuracy loss.

In contrast to adding components to the original DNN model (e.g., feature coding or early exit branches),
the 2-step-pruning framework proposed in this paper modifies and prunes the original DNN model.
Nevertheless these extra components can be added at the partition point after partitioning the DNN model between the mobile device and the edge server.
As a result, the proposed framework has good compatibility with existing feature coding and early exit techniques and can be applied conjunctively
with them to further improve the device-edge cooperative inference.

\section{Proposed Framework}

\begin{figure}[!t]
  \centering
  \includegraphics[width=3.2in]{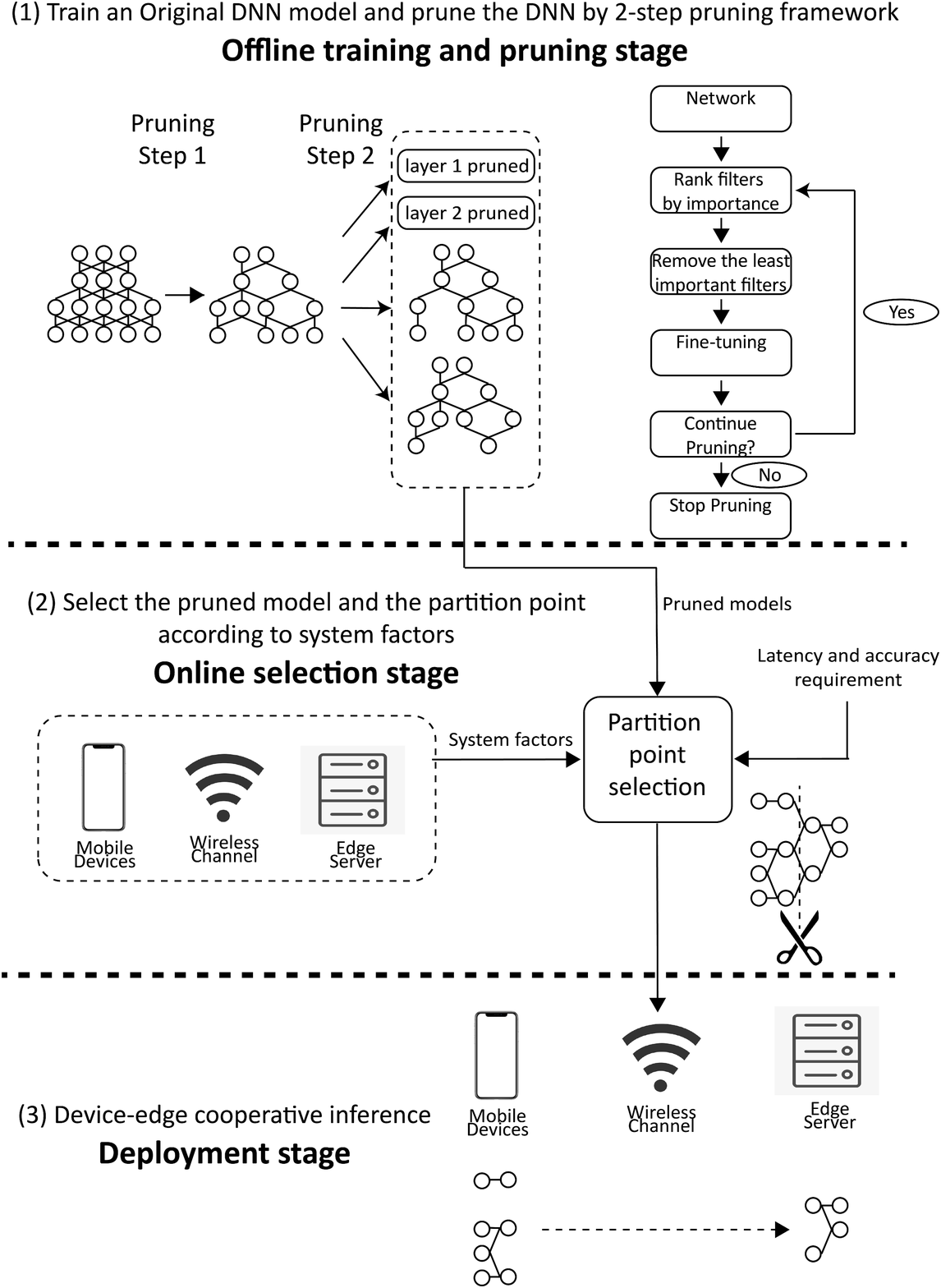}
  \caption{Proposed 2-step pruning framework.}
  \label{system}
  \end{figure}
In this section, we introduce our 2-step filter pruning framework to address the aforementioned challenges.
As shown in Fig.~\ref{system}, the proposed framework contains 3 stages: offline training and pruning stage, online partition point selection stage, and deployment stage.

\subsection{2-Step Filter Pruning in the Offline Training and Pruning Stage}
At the offline training and pruning stage, we use identical pruning workflow with NVIDIA\cite{molchanov2016pruning} and the iterative
pruning workflow used in both pruning steps is showned in the top right of Fig.~\ref{system}.
In the pruning workflow, a pruning range is first given according to whether it is pruning step 1 or step 2.
Then we rank all convolution filters in this range by their first order Taylor expansion on the network loss function and then remove insignificant ones.
After that we fine tune the pruned network and test its accuracy on the test dataset.
We continue the pruning iterations until the accuracy is lower than an given threshold.
Conventional pruning methods are designed to accelerate the computation of the DNN model.
However, reducing the transmission workload is also crucial to enhance the end-to-end latency in device-edge cooperative inference, thus we propose our 2-step pruning method.

In the \emph{\textbf{pruning step 1}}, the pruning range is the entire original testing network.
Experiments from our work and from reference work\cite{molchanov2016pruning} show that the testing neural network shrinks after this pruning step,
and the pruned network has a similar shape (i.e. filters number distribution of each layer) to the testing network.
As a result, the computation workload is reduced while the transmission latency still stays high when partitioning in the front end of the network.

The data needs to be transmitted via wireless channel is generated by the last layer in the front-end part of the network
before the partition point.
Based on this observation, we can prune the layer just before the partition point to shorten the transmission latency
after the partition point is selected.
But pruning will change the structure of the testing network which has a decisive impact on the partition point selection.
There is a tight coupling between partition point selection and pruning, 
and it is hard to decouple because of the difficulty in predicting how much filters can be pruned under a given accuracy constraint.
As a result, a brute force search process is of necessity.

In the \emph{\textbf{pruning step 2}}, 
we individually apply the pruning algorithm to each layer in the pruned network after pruning step 1 by restricting the pruning range to each layer.
A series of pruned network models are generated by pruning each layer, which correspond to the partition point just after the layer.
Hence, network models after pruning step 2 have only one different layer as compared to the network model after pruning step 1, as shown in the top left of Fig.~\ref{system}.
After the pruning is completed, all pruned models are profiled and stored for the next stage.

\subsection{Online Model and Partition Point Selection Stage}
In order to decide which pruned network model after pruning step 2 and its corresponding partition point has the lowest end-to-end latency under an given accuracy constraint,
some profiles are needed: (1) Layer level output data size and computation latency profiles of the pruned models, and these can be obtained from the testing in pruning step 2.
(2) The tolerable accuracy loss. 
(3) System factors such as the wireless channel condition, and the computation capabilities of the mobile device and the edge server.

For simplicity, we introduce two parameters: a computation capability ratio $\gamma$ and an average upload rate $R$ to illustrate system factors.
$\gamma$ is defined as
\begin{equation}
  \label{gamma}
  \gamma = \frac{t^{\rm{mobile}}_{i}}{t^{\rm{device}}_{i}},
\end{equation}
where $t^{\rm{mobile}}_{i}$ and $t^{\rm{device}}_{i}$ is the computation latency of the $i$th layer in the network on the mobile device and the edge server, respectively.
Further, we assume $\gamma$ remains the same for different $i$.
Since it is impossible for the mobile device to profile all pruned models due to the energy limitation, 
a $\gamma$ can be chosen according to the mobile device computation capability (e.g., CPU) and infer the mobile device side computation latency as
\begin{equation}
  \label{ratio}
  t^{\rm{mobile}}_{i} = \gamma \times t^{\rm{device}}_{i}.
\end{equation}

Additionally, an average upload rate $R$ is related to the wireless communication technique used in the system, 
and the transmission latency with partitioning at $i$th layer can be calculated as
\begin{equation}
  \label{rate}
  t^{\rm{transmission}}_{i} = \frac{D_i}{R},
\end{equation}
where $D_{i}$ is the volume of the $i$th layer output data which need to be transmitted.
Then an travelsal algorithm (algorithm~\ref{alg1}) can be used to find the best pruned DNN model and the corresponding partitioning point.

\begin{algorithm}
\caption{Pruned CNN and partition point seach}
\label{alg1}
\begin{algorithmic}[1]
\renewcommand{\algorithmicrequire}{\textbf{Input:}}
\renewcommand{\algorithmicensure}{\textbf{Output:}}
\REQUIRE \quad 
\\ $M$: number of pruned CNNs after the second pruning step (equal to layers number) 
\\ $ \{ L_i | i = 1, 2 \cdots, M \} $: layers in the CNN 
\\ $ \{ N_i | i = 1, 2 \cdots, M \} $: pruned CNNs 
\\ $ \{ A_i | i = 1, 2, \cdots, M \} $: accuracy of $N_i$ 
\\ $ \{ D_{i} | i = 1, 2 \cdots, M \} $: output data size for $L_i$ in $N_i$
\\ $ f(L_i)$:  accumulative computation latency upto $L_i$ in $N_i$ 
\\$T_i$: total computation latency of $N_i$ 
\\$\gamma$: the computation capability ratio of device and server 
\\$R$: average wireless upload rate 
\\$A$: allowed lowest accuracy
\ENSURE  \quad \\ Selection of the pruned CNN and the corresponding partition point
\FOR{ $i = 1,2 \cdots ,M $ }
  \IF{ $A_i > A$}
  \STATE {$t^{\rm{mobile}}_{i} \leftarrow \gamma \times f(L_i) $ }
  \STATE {$t^{\rm{device}}_{i} \leftarrow T_i-f(L_i)$}
  \STATE {$t^{\rm{transmission}}_{i} \leftarrow \frac{D_i}{R}$}
  \ENDIF
  \ENDFOR
  \IF {$ t^{\rm{mobile}}_{i}, \ t^{\rm{device}}_{i}, \ t^{\rm{transmission}}_{i}$ exist}
  \STATE $j = \mathop{\arg\min} \limits_{i=1,2,\cdots,M} (t^{\rm{mobile}}_{i}+t^{\rm{device}}_{i}+t^{\rm{transmission}}_{i})$
  \RETURN $L_j, N_j$
  \ELSE
  \RETURN NULL
  \ENDIF
\end{algorithmic} 
\end{algorithm}

\subsection{Deployment Stage}
At the deployment stage, the front end part of the selected pruned model is downloaded by the mobile device.
The mobile device can perform local computation up to the partition point and offload the rest to the edge server via the wireless channel.

\section{Experiments}

We use PyTorch, a deep learning framework in Python, in the following experiments.
Our server platform is shown in Tabel.~\ref{tab1}, GPUs are used in training and pruning while CPU is used in profiling the pruned DNN models.
VGG\cite{simonyan2014very}, a state-of-the-art CNN for image classification, is the target network for device-edge cooperative inference.
Our dataset is CIFAR-10\cite{krizhevsky2014cifar}, a widely used image classification dataset with 10 classes of objects.
Typical average \emph{upload} rate $R$ of 3G, 4G and WiFi network is 1.1Mbps, 5.85Mbps and 18.88Mbps, respectively\cite{state, state2}.
We range $\gamma$ between 0.1 and 100 to simulate various computation capabilities of mobile devices.
Since the tolerable accuracy loss of the two pruning steps is related with the application requirements,
we adopt 4\% total accuracy loss threshold in the experiments for general cases.

\begin{table}[htbp]
\caption{Server Platform Specifications}
\label{tab1}
\begin{center}
\begin{tabular}{|c|c|}
\hline
\textbf{Hardware} & \textbf{Specifications} \\
\hline
System & Supermicro SYS-7048GR-TR, 4$\times$PCIe 3.0 $\times$16 slots\\
\hline
CPU & 2 $\times$ Intel Xeon E5-2640 V4, 2.4GHz \\
\hline
Memory & 128GB DDR4 2400MHz \\
\hline
GPU & 4 $\times$ NVIDIA TITAN Xp \\
\hline
\end{tabular}
\end{center}
\end{table}

\subsection{End-to-End Latency}

\begin{figure}[!t]
  \centering
  \includegraphics[width=3.2in]{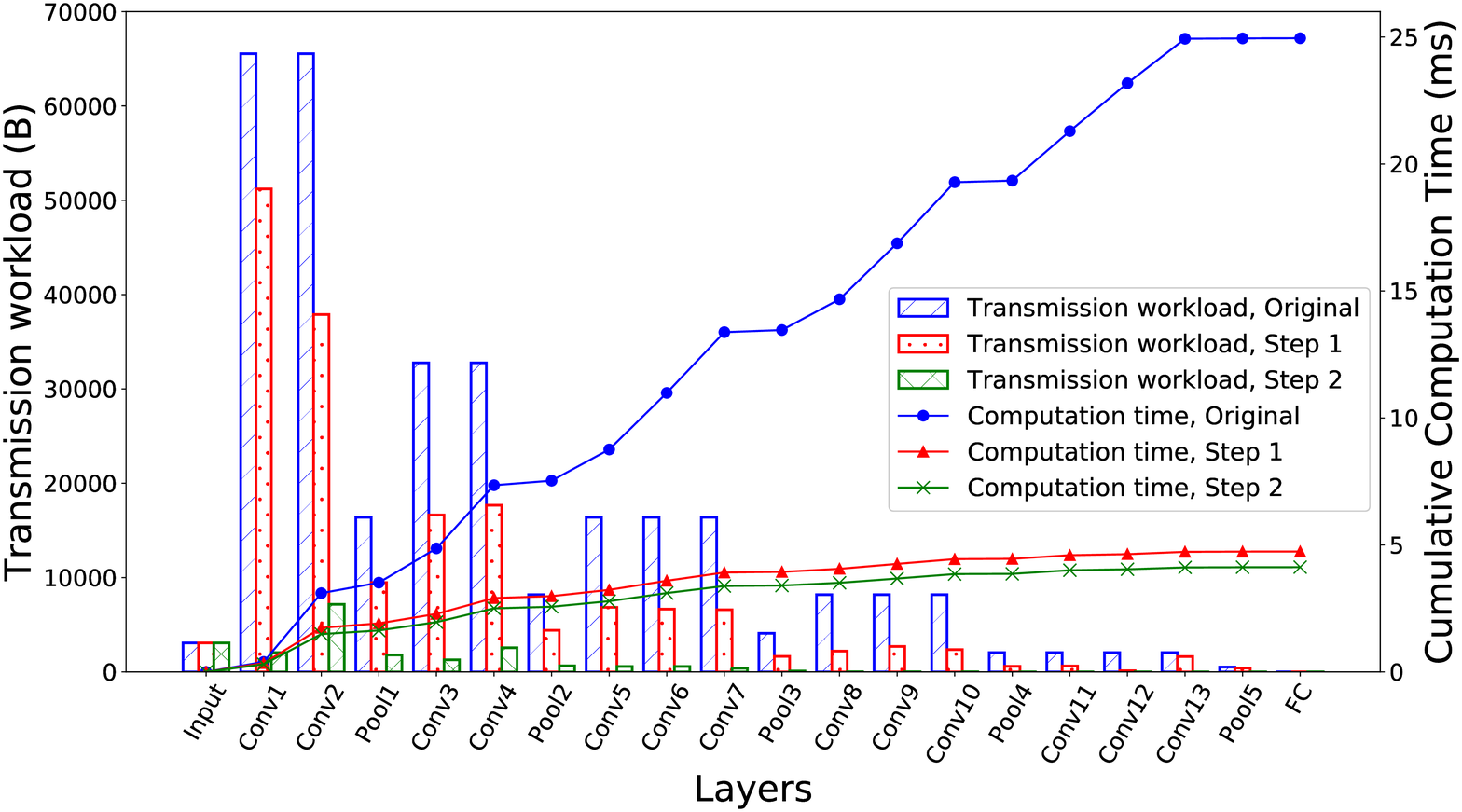}
  \caption{Layer level transmission and computation characteristics of the original, step 1 pruned and step 2 pruned VGG.}
  \label{fg2}
\end{figure}

\begin{figure*}[!t]
  \centering
  \subfloat[]{\includegraphics[width=2.1in]{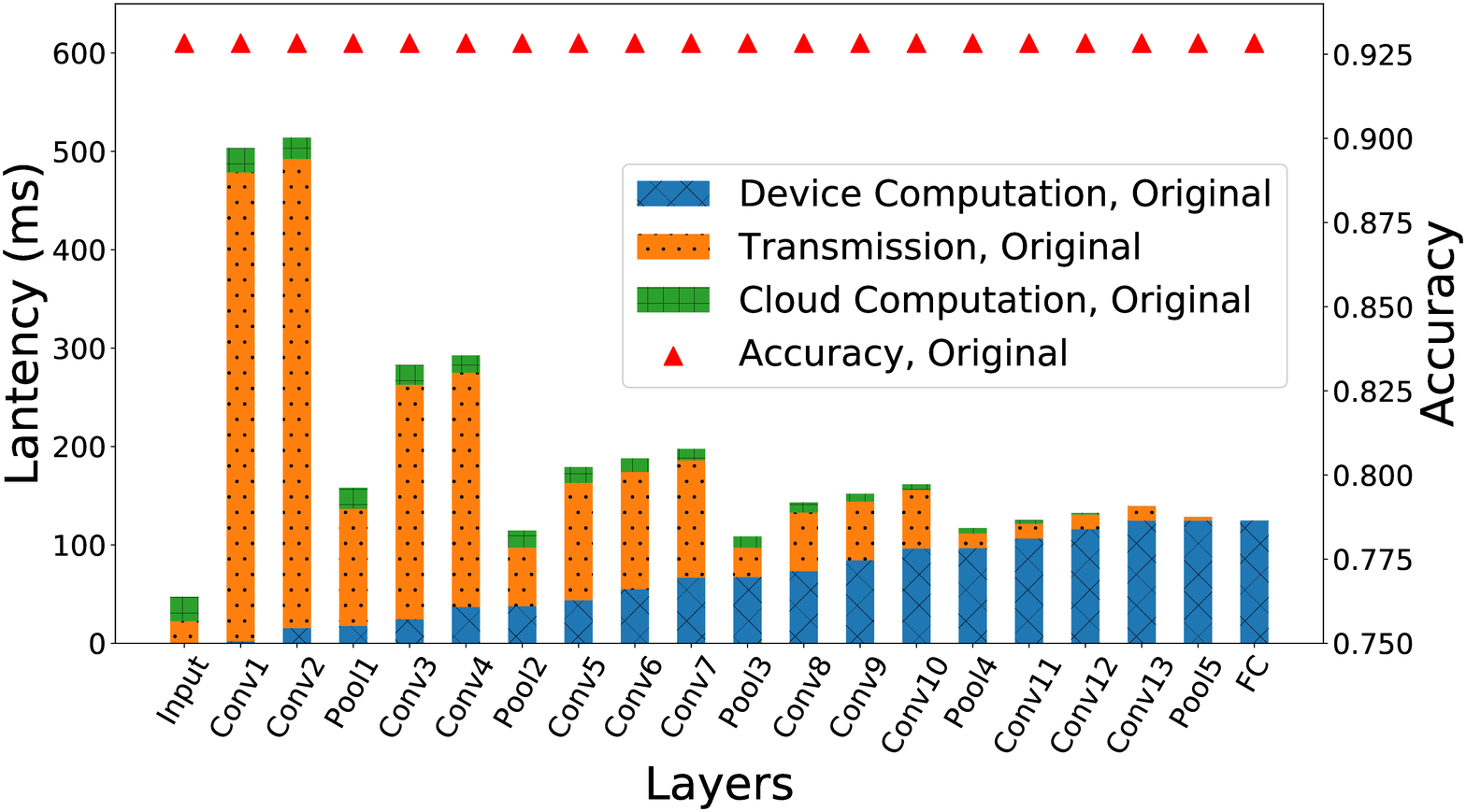}
  \label{Original VGG}}
  \hfil
  \subfloat[]{\includegraphics[width=2.1in]{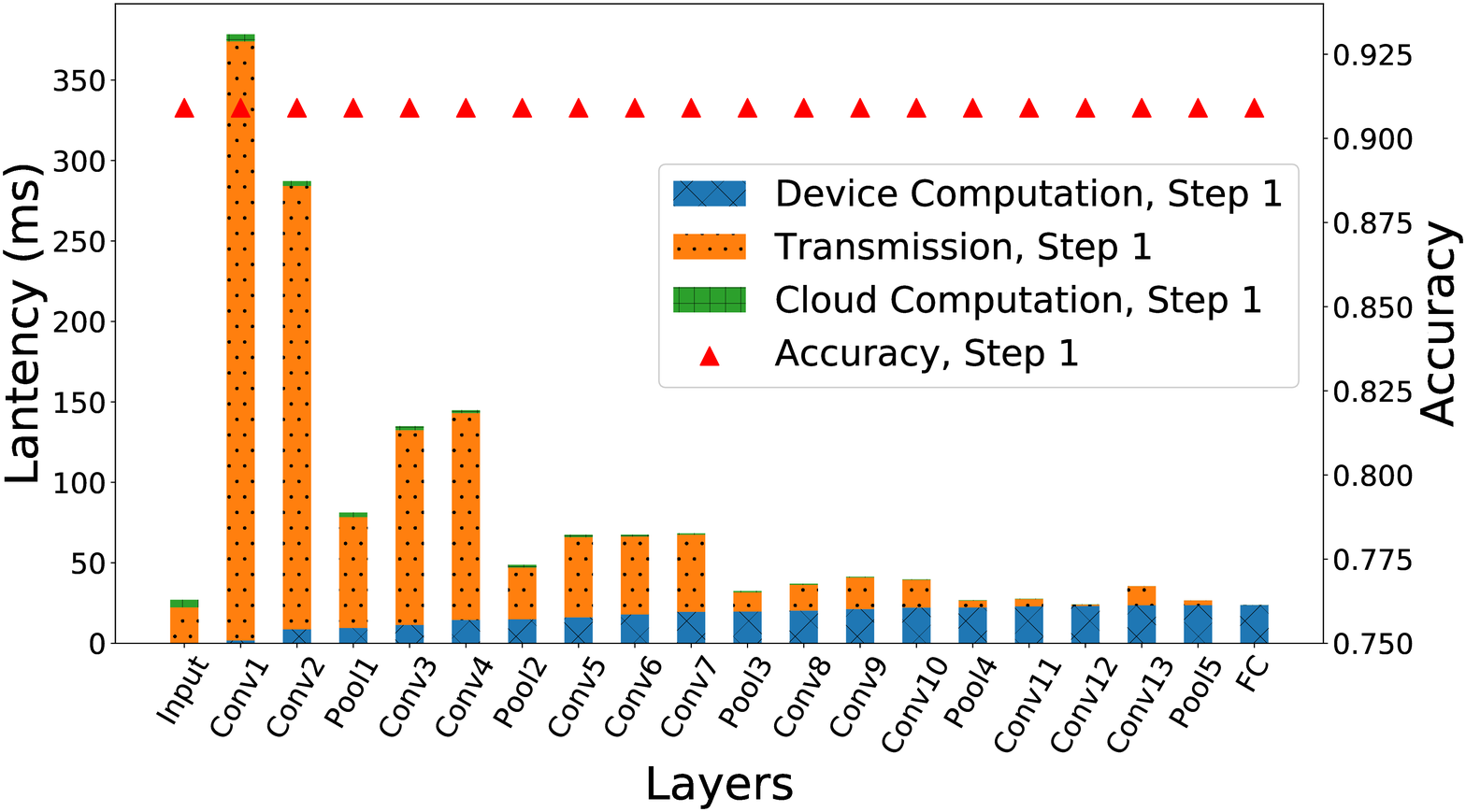}
  \label{Pruned VGG after pruning step 1}}
  \hfil
  \subfloat[]{\includegraphics[width=2.1in]{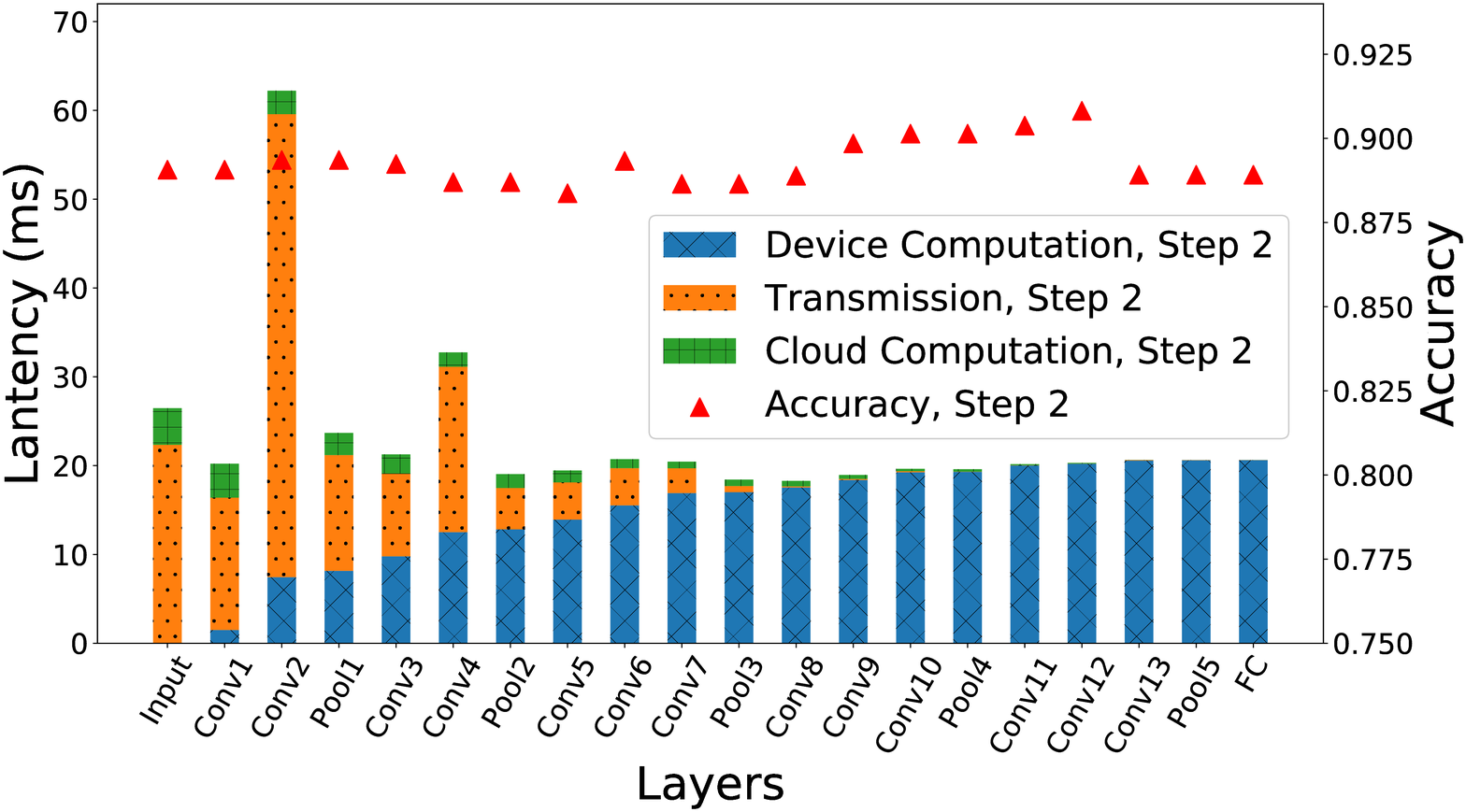}
  \label{Pruned VGG after pruning step 2}}
  \caption{End-to-end latency and accuracy vs. the partitoin point of the original, step 1 pruned and step 2 pruned VGG, respectively. 
  The average upload rate is $R=137.5$KB/s and the computation capability ratio is $\gamma=5$.}
  \label{fg3}
\end{figure*}    

\begin{figure*}[!t]
  \centering
  \subfloat[]{\includegraphics[width=1.6in]{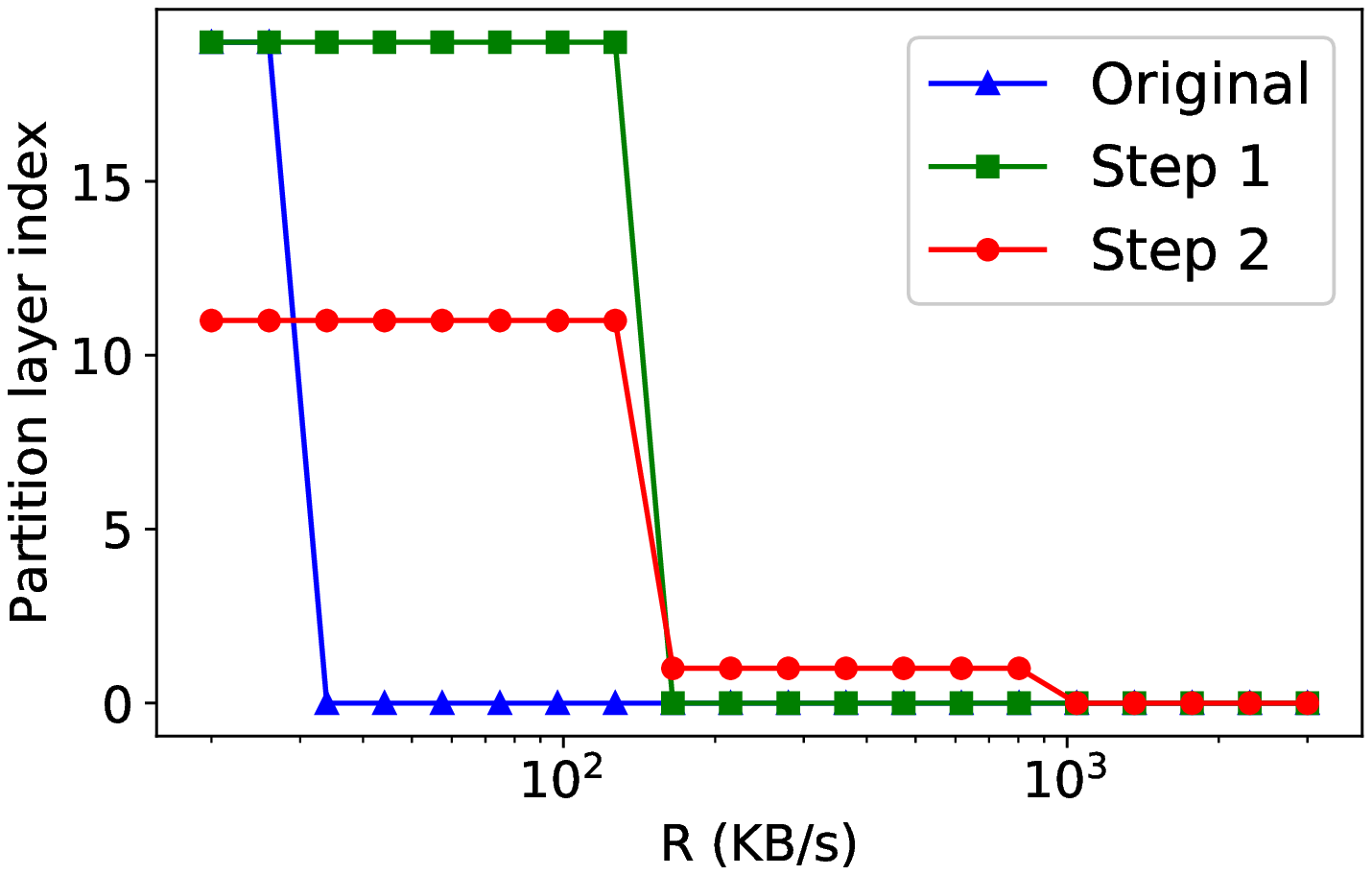}
  \label{partition_over_R}}
  \hfil
  \subfloat[]{\includegraphics[width=1.6in]{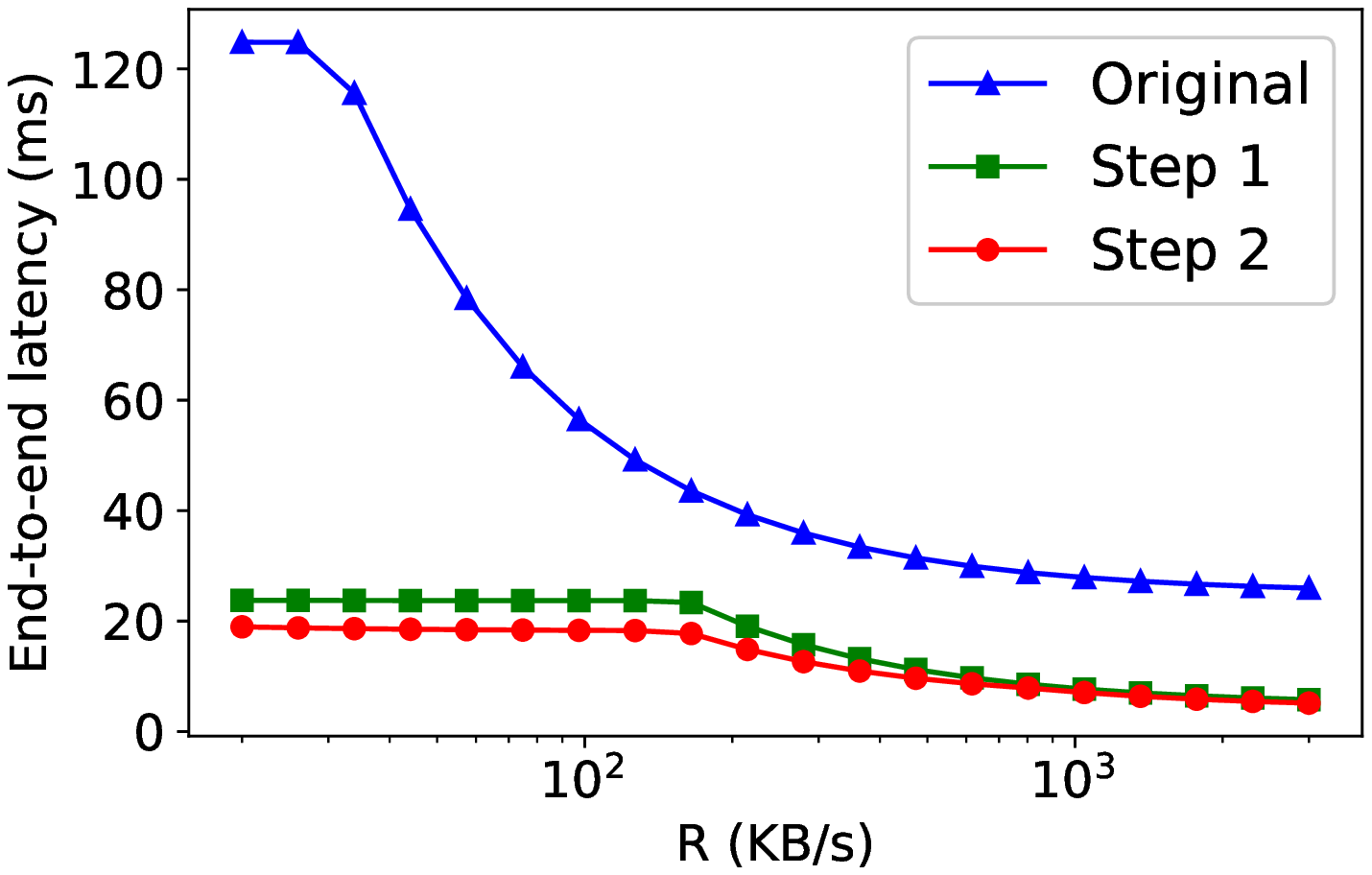}
  \label{latency_over_R}}
  \hfil
  \subfloat[]{\includegraphics[width=1.6in]{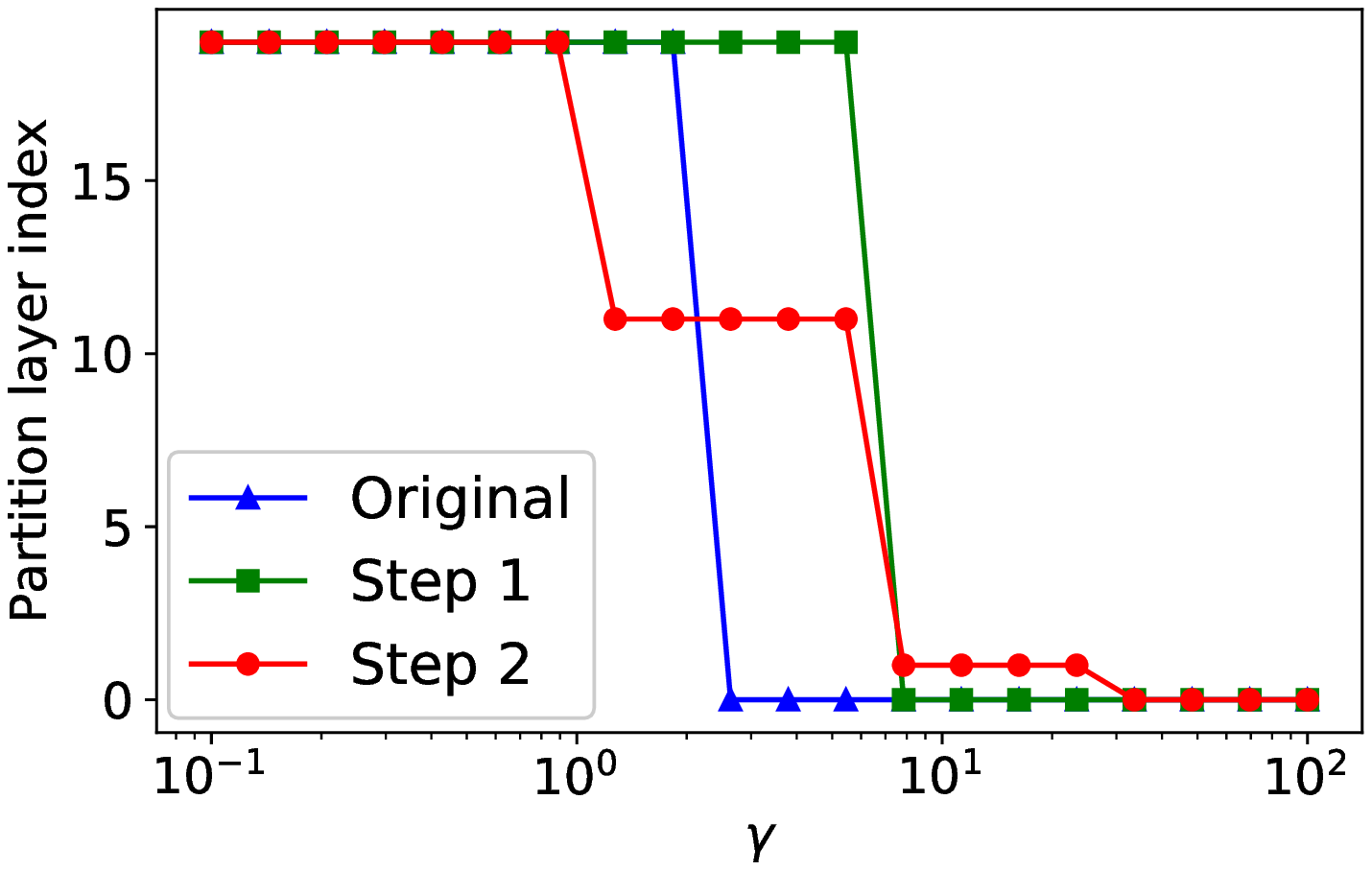}
  \label{partition_over_gamma}}
  \hfil
  \subfloat[]{\includegraphics[width=1.6in]{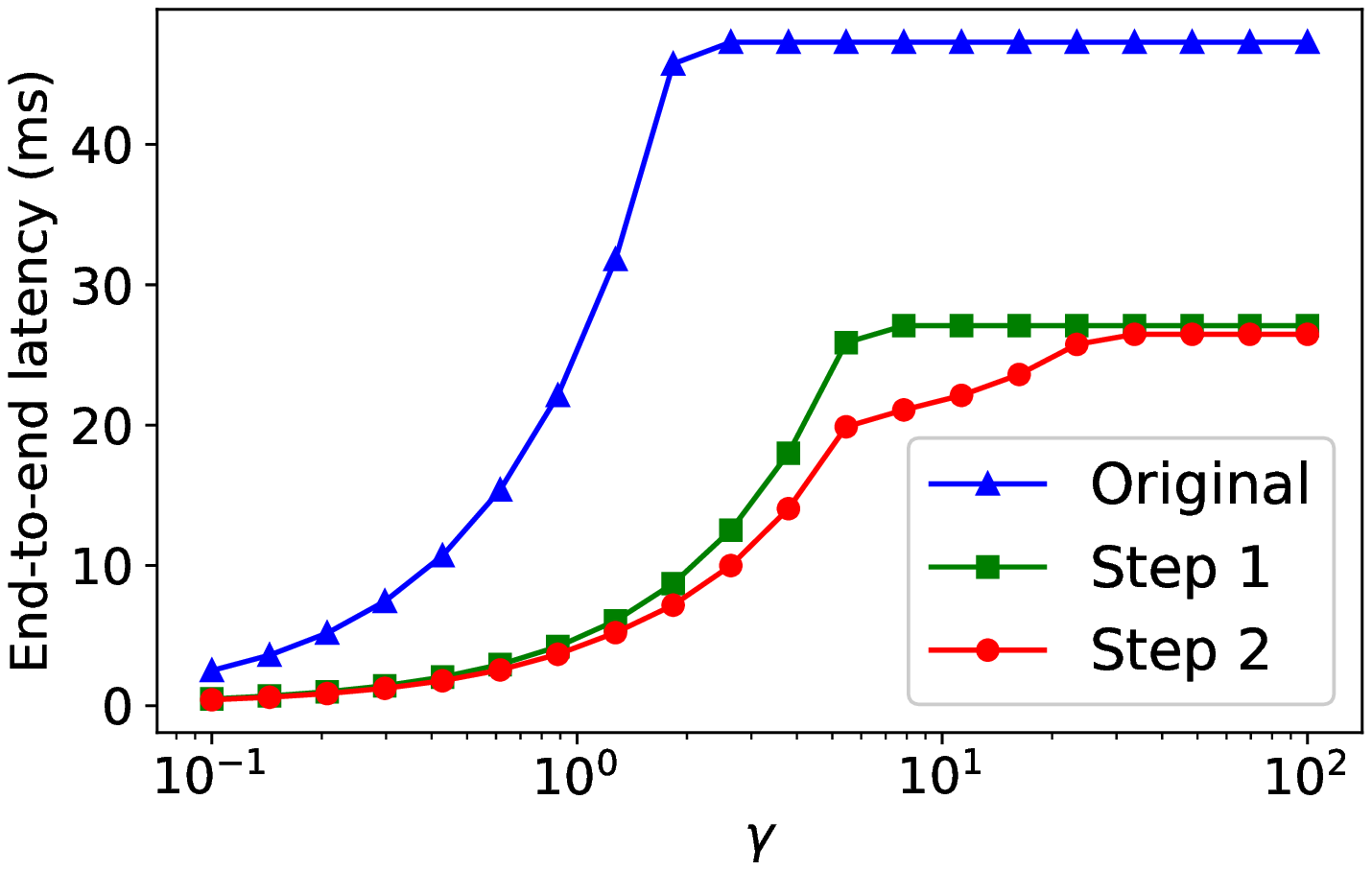}
  \label{latency_over_gamma}}
  \caption{Partition point selection and end-to-end latency performance vs. system factors.
  (a) and (b) show the partition point and corresponding end-to-end latency vs. average upload rate $R$.
  (c) and (d) show the partition point and corresponding end-to-end latency vs. computation capability raito $\gamma$.}
  \label{fg4}
\end{figure*}

In this section, we examine the end-to-end latency by applying the proposed 2-step pruning framework.
Fig.~\ref{fg2} shows the transmission workload and cumulative computation time at each layer for the original VGG model, 
VGG model after pruning step 1 and VGG model after pruning step 2, respectively.
The left bars in the histogram in Fig.~\ref{fg2} show the transmission workload in the original VGG at each layer. 
Due to the increasing number of feature maps at the front-end part of the network, the intermediate data get an order of magnitude larger than the input, 
which will bring large transmission latency if we partition the original VGG in the front-end part of the network. 
The middle bars in the histogram are for the pruned VGG model after pruning step 1. 
Since filters in the back-end part of the network get a higher probability to be pruned due to the pruning algorithm we used\cite{molchanov2016pruning},
partitioning in the front-end of the network will face even severe transmission latency issue (Fig.~\ref{fg3}\subref{Pruned VGG after pruning step 1}).
Nevertheless, the curves show that using step 1 pruning can reduce the overall computation time up to 5.35$\times$.

The right bars in the histogram are for the pruned VGG model after pruning step 2.
Remark that each bar stands for a specific pruned VGG. 
For example, the right bar with index ``conv1'' presents a model with ``conv1'' layer pruned by pruning step 2 and other layers are equivalent to the pruned VGG after pruning step 1.
Since the pruning step 2 only prunes one layer in the network, it can significantly reduce the transmission workload but with only a little computation latency reduction.
Combining two pruning steps, we can get up to 25.6 $ \times $ reduction in transmission workload and 6.01 $\times$ acceleration in computation compared to the original model.

Then we use an average upload transmission rate $R=137.5$kB/s (3G network) and a computation capability ratio $\gamma = 5$ as a baseline system configuration.
Fig.~\ref{fg3} shows the end-to-end latency and accuracy in this setting.
The histograms in the figure are the end-to-end latency with partitioning in each layer, and we use colors to distinguish different latency components.
Fig.~\ref{fg3}\subref{Pruned VGG after pruning step 2} shows that the accuracy loss due to the step 2 pruning varys over different layers.
If the system allows low-accuracy (e.g., 88\% accuracy), then using partition point ``Pool 3'' can achieve the best latency performance.
When a precise network (e.g., 90\% accuracy) is needed, the partition point should be moved to the ``Pool 4''.

\subsection{Partition Point Selection}

In this section, we examine the best partition point under various system configurations.
Fig~\ref{fg4}\subref{partition_over_R} shows the partition points identified by the selection algorithm under various average uplink rates with $\gamma=5$, 
and Fig~\ref{fg4}\subref{partition_over_gamma} shows the results under various computation capability ratios with $R=137.5$kB/s, respectively. 
Fig~\ref{fg4}\subref{latency_over_R} and Fig~\ref{fg4}\subref{latency_over_gamma} show the corresponding end-to-end latency. 
Due to the high transmission workload for the intermediate data, 
the original NN prefers completing the entire inference either at local device (partition layer index is 0) or at edge server (partition layer index is 18). 
The step 1 pruned NN is similar while it has relatively less computation workload and prefers local computation to edge computation in bad transmission 
conditions or under poor edge computation capability settings. 
However, the system can benefit from partitioning after step 2 pruning.
Table.~\ref{tab3} shows the end-to-end latency improvements for 3 typical mobile networks (3G, 4G and WiFi) with $\gamma=5$.
By applying 2-step-pruning framework, 4.81$\times$ acceleration can be achieved in WiFi environments.

\begin{table}[!t]
  \caption{End-to-End Latency Improvements under 3 Typical Mobile Networks with $\gamma=5$}
  \label{tab3}
  \begin{center}
  \begin{tabular}{|c|c|c|c|}
  \hline
  \textbf{Network} &\textbf{Original (ms)} &\textbf{Step 2 purning (ms)} &\textbf{Improvement} \\
  \hline
  3G & 46.64 & 17.84 & 2.61 $\times$\\
  \hline
  4G & 28.50 & 7.73 &  3.69 $\times$\\
  \hline
  WiFi & 25.61 & 5.32 & 4.81 $\times$\\
  \hline
  \end{tabular}
  \end{center}
\end{table}

\subsection{Bandwidth-Accuracy Tradeoffs}

Since the max pooling layers reduce the data volume by 4$\times$ without introducing much extra computation cost, 
it is reasonable to partition the network at a max pooling layer which is also confirmed by Fig.~\ref{fg4}\subref{partition_over_R} and Fig.~\ref{fg4}\subref{partition_over_gamma}
In this section, all investigations are based on the assumption that we parttion the network at one of the max pooling layers.

\begin{figure*}[!t]
  \centering
  \subfloat[]{\includegraphics[width=2.2in]{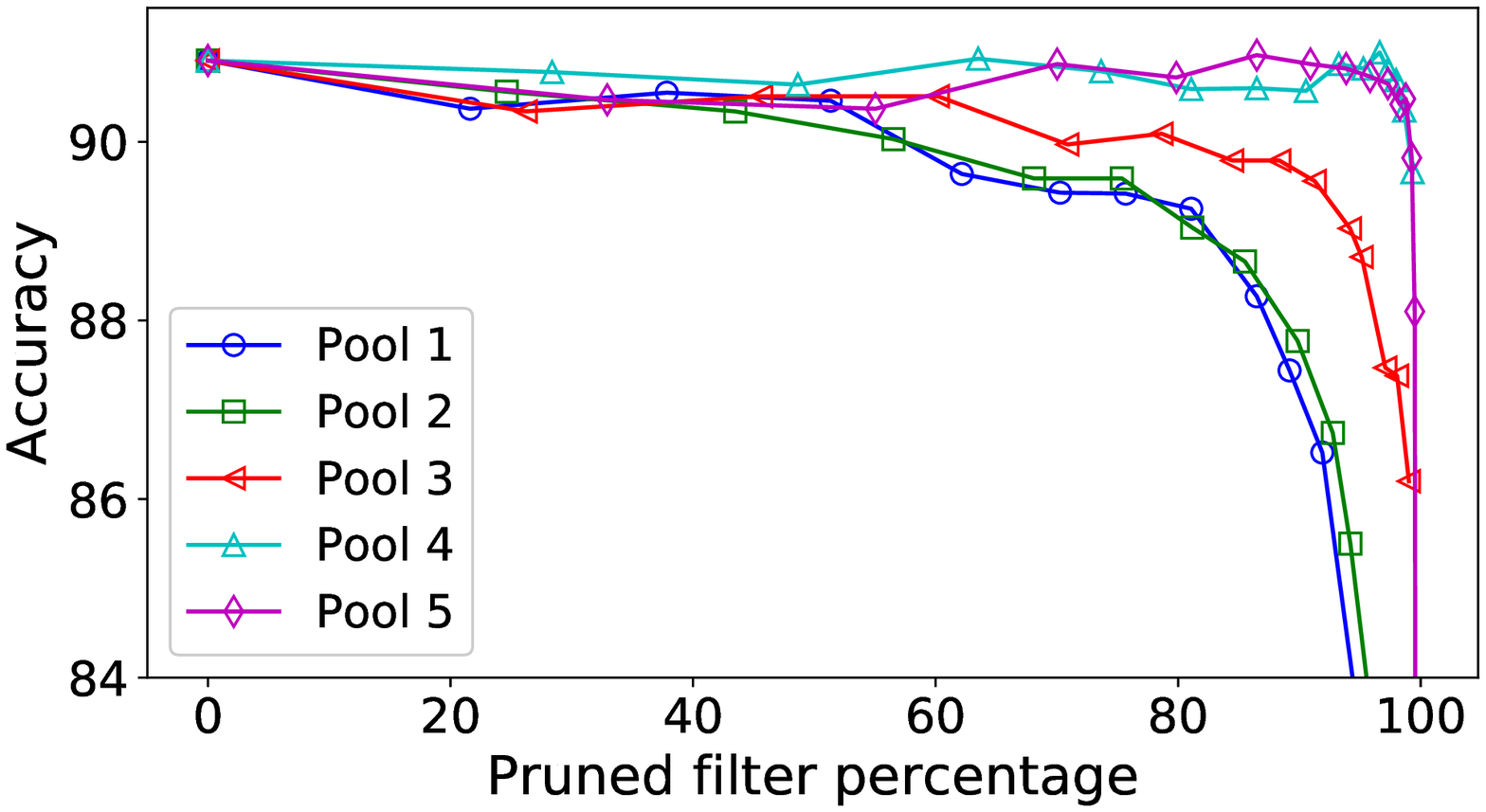}
  \label{percentage}}
  \hfil
  \subfloat[]{\includegraphics[width=2.2in]{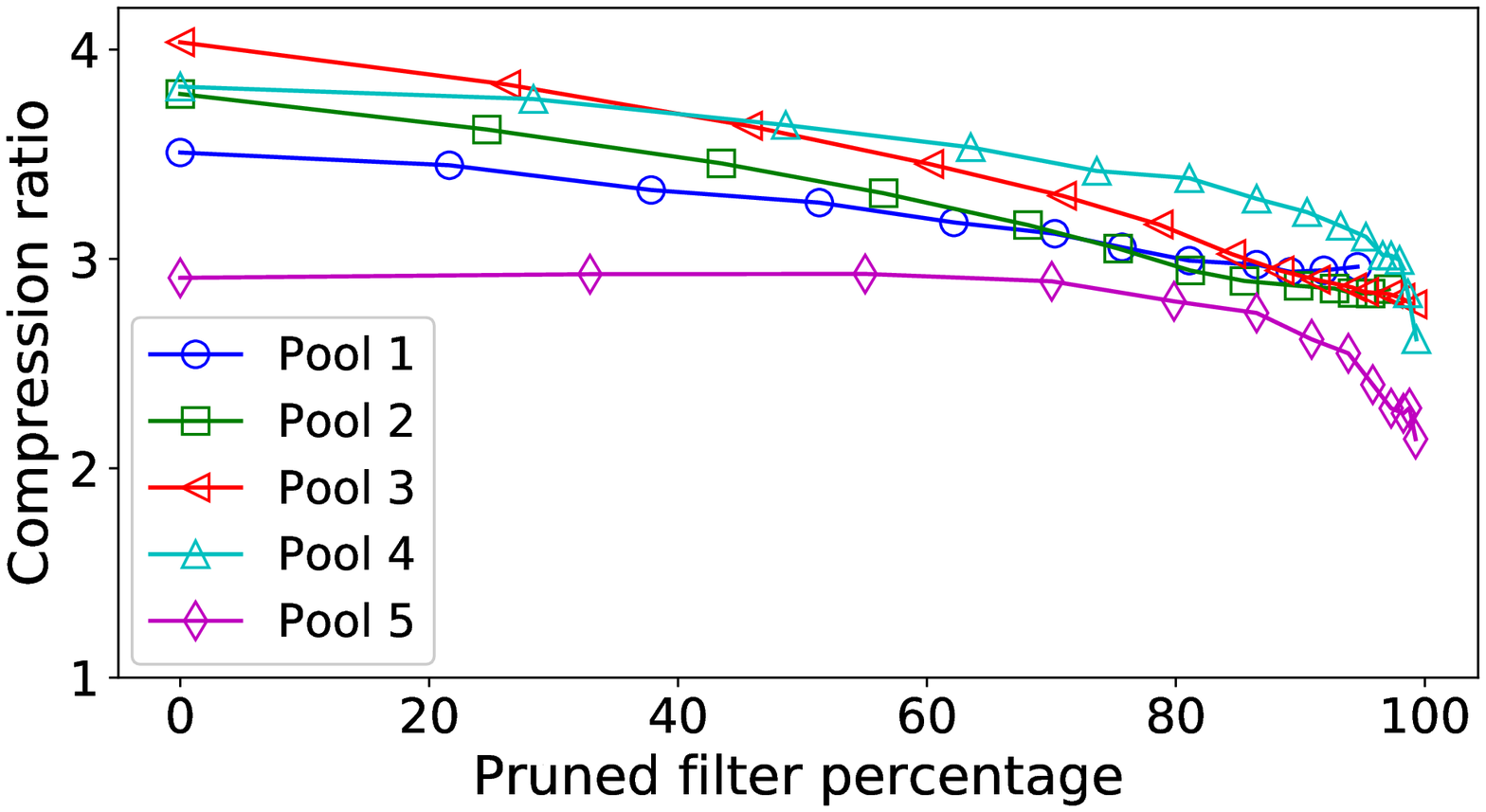}
  \label{PNG}}
  \hfil
  \subfloat[]{\includegraphics[width=2.2in]{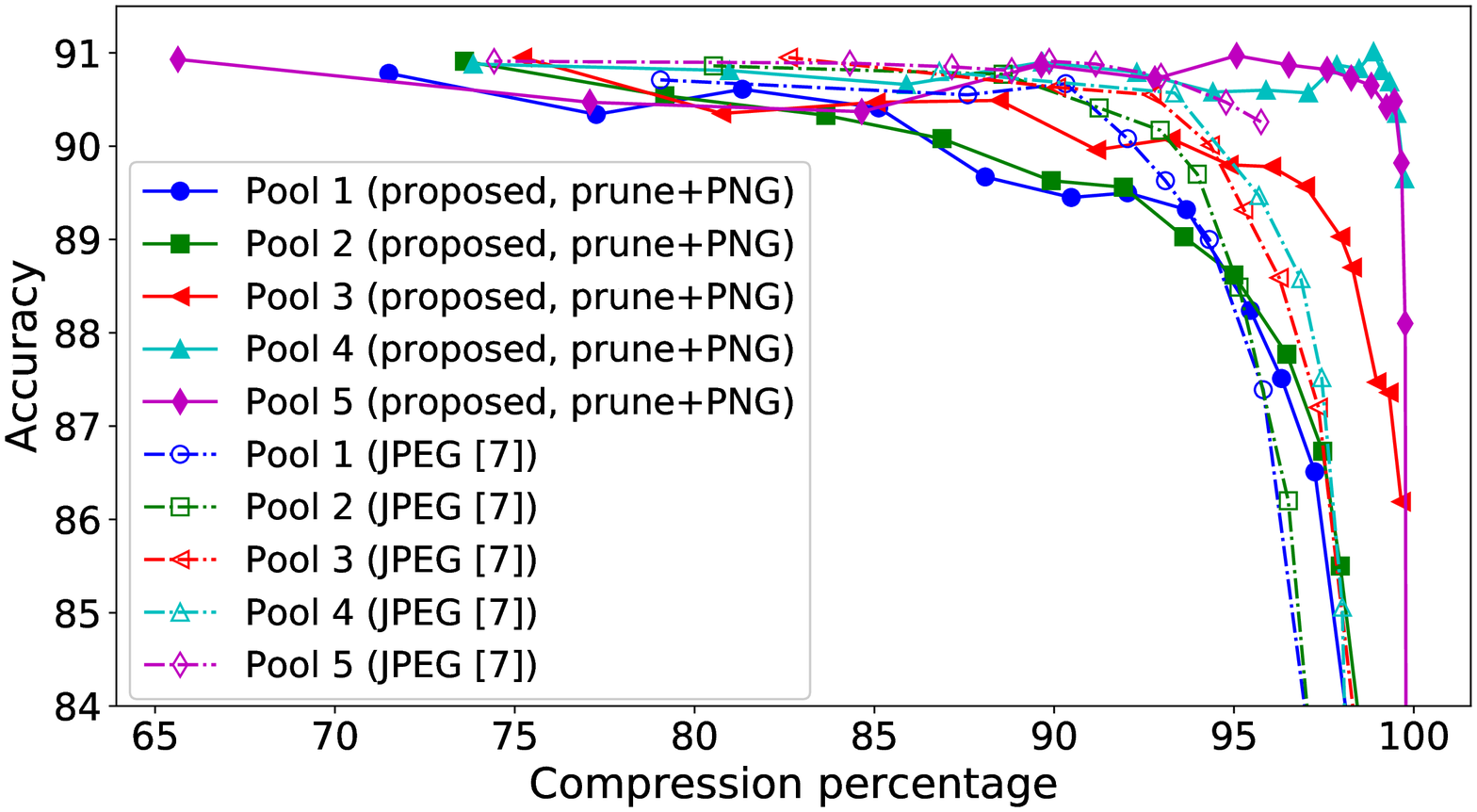}

  \label{prune+png_jpeg}}
  \caption{Transmission workload and accuracy tradeoffs. 
  (a) shows the prune-accuray tradeoff for the step 2 pruning. 
  (b) shows the extra compression effects by adding PNG coding on the intermediate data for transmission in the 2-step pruning framework.
  (c) shows the comparision of the pruposed pruning + PNG approach with JPEG coding [7] on the prune-accuracy tradeoff.}
  \label{fg5}
\end{figure*}   

The proposed framework generates a series of pruned DNN models for each max pooling layer after the offline training and pruning stage, 
then selects a proper pruned DNN model to satisfy accuracy and latency constraints.
Fig.~\ref{fg5}\subref{percentage} shows the tradeoff between the accuracy and pruned filter percentage after pruning step 2.
As shown in Fig.~\ref{fg5}\subref{percentage}, the layers in the front end part of the network (e.g., Max pool 1 and Max pool 2) 
are more sensitive to pruning compared with the layers in the back end.

In our framework, extra compression component can be added before the partition point to provide further compression performance improvement.
Fig.~\ref{fg5}\subref{PNG} shows the extra compression effect provided by adding an lossless PNG encoder and decoder at each max-pooling layer (as potential partition point)
of the pruned VGG models after pruning step 2.
The extra compression provided by PNG encoder decreases as the percentage of the pruned filter increases, 
since pruning more filters corresponds to less redundancy in the intermediate data.
Nevertheless, the lossless PNG encoder and decoder can compress the intermediate data at least 2$\times$ without any accuracy loss.

Finally we compare the proposed 2-step pruning framework with the feature coding\cite{ko2018edge}.
In the feature coding approach, standard JPEG encoding is used to compress the feature maps at the partition point of the VGG model generated by pruning step 1, 
and then fine-tune the rest of the VGG model after JPEG decoding to eliminate the accuracy loss introduced by JPEG coding.
Fig.~\ref{fg5}\subref{prune+png_jpeg} compares the transmission workload reduction between the feature coding
and 2-step pruning with PNG coding. 
It is shown that the proposed approach outperforms the feature coding approach\cite{ko2018edge} when partitioning at the back-end part of the network.
While in the front-end part of the network, 2-step-pruning performs better in the high compression ratio region.

\section{Conclusion}  
In this paper, we propose an efficient and flexible 2-step pruning framework.
In the proposed framework, the mobile devices can complete a computation intensive CNN inference under latency 
and accuracy constraints by offloading part of the computation to the edge servers.
Our framework contains 3 stages: training and pruning stage, selection stage and deployment stage.
Our simulations show that pruning step 1 can greatly reduce the total computation workload of the network while pruning step 2 can effectively compress the intermediate data
for transmission. 
The five max pooling layers in the testing network (VGG) are the most likely partition points.
Combining two pruning steps can achieve 25.6 $ \times $ reduction in transmission workload and 6.01 $ \times $ acceleration in computation with less than 4\% accuracy loss.
By properly choosing the partition point, up to 4.81$\times$ end-to-end latency improvement can be achieved under WiFi environment.
The proposed 2-step-pruning framework is compatible with existing feature coding techniques, e.g., PNG coding.
Our approach can maintain higher accuracy than existing lossy feature coding techniques under the bandwidth constraint in most circumstance.

\section*{Acknowledgement}
This work is sponsored in part by the Nature Science Foundation of China
 (No. 61871254, No. 91638204, No. 61571265, No. 61861136003, No. 61621091), National Key R\&D Program of China 2018YFB0105005, and Hitachi Ltd.

\bibliographystyle{IEEEtran}
\bibliography{reference}

\end{document}